\documentclass[letterpaper, 10 pt, conference]{ieeeconf} 
\IEEEoverridecommandlockouts                              
\usepackage{graphicx}
\usepackage{dcolumn}
\usepackage{bm}
\usepackage{mathbbol}

\usepackage{amssymb}          
\usepackage{graphicx}         
\usepackage[latin1]{inputenc} 
\usepackage{amsmath}          
\usepackage{color}
\usepackage{mathabx}
\usepackage{epstopdf}

\newtheorem{prop}{Proposition}
\newtheorem{defin}{Definition}
\newtheorem{thm}{Theorem}
\newtheorem{cor}{Corollary}
\newtheorem{lemma}{Lemma}

\renewcommand{\proof}{\noindent {\em Proof. }}

\newcommand{\ket}[1]{|#1\rangle}
\newcommand{\bra}[1]{\langle #1|}
\newcommand{\ketbra}[1]{|#1\rangle\langle #1|}

\newcommand{\Hi}{\mathcal{H}}

\newcommand{\Ei}{\mathcal{E}}
\newcommand{\Ec}{\mathcal{E}}

\newcommand{\Tr}{\mathrm{{Tr}}}
\newcommand{\supp}{\textrm{supp}}

\newcommand{\Ni}{\mathcal{N}}
\newcommand{\neigh}{\mathcal{N}}
\newcommand{\fix}{\textrm{Fix}}
\newcommand{\identity}{\mathbb{I}}

\newcommand{\beq}{\begin{equation}}
\newcommand{\eeq}{\end{equation}}
\newcommand{\beqa}{\begin{eqnarray}}
\newcommand{\eeqa}{\end{eqnarray}}
\newcommand{\beqan}{\begin{eqnarray*}}
\newcommand{\eeqan}{\end{eqnarray*}}

\newcommand{\qed}{\hfill $\Box$ \vskip 2ex}

\renewcommand{\span}{\rm{span}}

\title{\LARGE {\bf Distributed finite-time stabilization of entangled quantum states\\ on tree-like hypergraphs}*}

\author{Francesco Ticozzi, Peter D. Johnson, and Lorenza Viola
\thanks{*This work was partially supported by the QFUTURE and QCOS projects of the Universit\`a 
degli Studi di Padova, and by the National Science Foundation grant No. PHY-1620541.}
\thanks{F. Ticozzi is with the Dipartimento di Ingegneria
dell'Informazione, Universit\`a di Padova, via Gradenigo 6/B,
35131 Padova, Italy, and the Department of Physics and Astronomy, Dartmouth College, 6127 Wilder Laboratory, 
Hanover, NH 03755, USA 
({\tt ticozzi@dei.unipd.it}).}
\thanks{P. D. Johnson is with the Department of Chemistry and Chemical Biology, 
Harvard University, 12 Oxford Street, Cambridge, MA 02138, USA ({\tt peter.d.johnson22@gmail.com}).}
\thanks{L. Viola is with the Department of Physics and Astronomy, Dartmouth College, 6127 Wilder Laboratory, 
Hanover, NH 03755, USA ({\tt lorenza.viola@dartmouth.edu}).}
}

\begin{document}
\maketitle

\begin{abstract}
Preparation of pure states on networks of quantum systems by controlled dissipative dynamics offers important advantages 
with respect to circuit-based schemes.
Unlike in continuous-time scenarios, when discrete-time dynamics are considered, dead-beat stabilization 
becomes possible in principle. Here, we focus on pure states that can be stabilized by distributed, unsupervised dynamics in finite time 
on a network of quantum systems subject to realistic quasi-locality constraints. In particular, we define a class of quasi-locality notions,
that we name ``tree-like hypergraphs,'' and show that the states that are robustly stabilizable in finite time 
are then unique ground states of a frustration-free, commuting quasi-local Hamiltonian. A structural characterization of such 
states is also provided, building on a simple yet relevant example.

\end{abstract}

\section{Introduction}
Recent efforts aimed to develop viable quantum technologies have increasingly recognized that access to controlled 
dissipative dynamics may offer distinctive advantages and unique capabilities across a variety of quantum tasks, ranging 
from universal open-system engineering and dissipation-driven computation \cite{Viola2001,Cirac2009} to 
sequential generation \cite{schon2005} and stabilization of entangled states of interest \cite{kraus-dissipative}. 
Experimental proof-of-principle demonstrations of dissipative entangled-state preparation have been reported for platforms 
as diverse as atomic ensembles \cite{Cirac2011}, trapped ions \cite{Barreiro-Nature:11,Yin2013}, superconducting qubits 
\cite{Devoret,Siddiqi} and NV centres in diamond \cite{Wrachtrup2016}.

In view of the above progress, it becomes important to devise stabilization schemes for multipartite entangled states,  
that can accommodate realistic resource constraints and, ideally, allow for scalable and robust implementation -- 
for instance, using distributed, possibly randomized dynamics. 
A system-theoretic approach to characterize stabilizable states under locality-constrained dynamical semigroups has been proposed in \cite{Ticozzi2012,TicozziQIC2014,Johnson2016,heisenberglyap}, and has been recently extended to {\em discrete-time Markov dynamics} 
in \cite{Ticozzi-alternating,johnson-FTS}.  Beside lending itself naturally to describe 
{\em digital} open-system simulators \cite{Blatt2013, Liang2016}, 
the discrete-time setting and the alternating-projection methods of \cite{Ticozzi-alternating} 
open the door to achieve finite-time stabilization, namely, to reach an 
invariant target state with {\em zero error in finite time}. This cannot be done with continuous-time Markov dynamics, 
even allowing for time-inhomogenous evolution \cite{johnson-FTS}.
Sufficient conditions for the existence of sequences of locality-constrained quantum maps ensuring finite-time stabilization of a target 
pure state are provided in \cite{johnson-FTS}. When, in addition, robustness with respect to the order of the maps in the sequence is
demanded, they always imply the existence of a Hamiltonian that is the sum of {\em commuting} quasi-local components, and for 
which the target is the unique ground state.

In general, whether the latter is actually necessary for robust finite-time stabilization remains an open question. 
Here, by specializing to a relevant class of quasi-locality notions, we are able to provide a full characterization of 
finite-time robustly stabilizable states, and a positive answer to the above question.
The locality notion we study is associated to hypergraphs that have a tree-like structure, 
as we will formally define in Section \ref{ssec:treelike}, and includes as particular cases linear graphs and trees.

\section{Preliminaries and existing results}

\subsection{Quasi-locality constraints via hypergraphs}

We shall focus on a finite-dimensional, multipartite quantum system
consisting of $n$ (distinguishable) subsystems, defined on a tensor-product 
Hilbert space,
$$\Hi=\bigotimes_{a=1}^n\Hi_a, \;\;\; \;a=1,\ldots,n,\;\; 
\text{dim}(\Hi_a) \equiv d_a<+\infty. $$
The state of the system is described by a density operator $\rho\in{\cal D}(\Hi),$ 
where ${\cal D}(\Hi) \subset {\cal B}(\Hi)$ denotes the set of trace-one, positive semidefinite operators 
in the set of all linear operators ${\cal B}(\Hi)$ on $\Hi$.

In order to account for physical {locality constraints} on operators, measurements, and dynamics on $\Hi,$ we impose 
a {\em neighborhood structure} on $\Hi$. Following \cite{Ticozzi2012}, neighborhoods ${\cal N}_j$ are 
subsets of indexes labeling the subsystems, that is,
\({\cal N}_j\subsetneq\{1,\ldots,n\}, \quad j=1,\ldots, N.\)
Mathematically, neighborhoods are hyperedges specifyng an {\em hypergraph} \cite{hypergraph}, 
which we refer to as a neighborhood structure, ${\cal N}\equiv \{ {\cal N}_j\}$.
A {\em neighborhood operator} $M$ is an operator on $\Hi$ such that there exists a neighborhood ${\cal N}_j$ 
for which we can write 
\(M \equiv M_{{\cal N}_j}\otimes I_{\overline{\cal N}_j},\)
where $M_{{\cal N}_j}$ accounts for the action of $M$ on subsystems in ${\cal N}_j$, and 
$I_{\overline{\cal N}_j}\equiv \bigotimes_{a\notin{\cal N}_j}I_a$ is the identity on the
remaining ones.
\noindent 
Once a state $\rho$ and a neighborhood structure are assigned on $\Hi,$ a list of 
{\em neighborhood reduced states} can be computed by letting 
\begin{equation} 
\label{redstate} 
\rho_{{\cal N}_j} \equiv \mbox{Tr}_{\overline{\cal N}_j}(\rho),
\quad \rho \in {\mathfrak D}(\Hi), \;\; j=1,\ldots, N,
\end{equation}
where $\mbox{Tr}_{\overline{\cal N}_j}$ indicates the partial trace over the 
tensor complement of ${\cal N}_j,$ namely,
$\Hi_{\overline{\cal N}_j} \equiv \bigotimes_{a\notin{\cal N}_j}\Hi_a$.  

\subsection{Quasi-local Markov dynamics and stabilization}
We consider general non-homogeneous, discrete-time Markov dynamics on $\Hi.$ In the quantum domain, 
the role of stochastic matrices is taken by completely-positive (CP), trace-preserving (TP) quantum maps. 
A CP map is a linear map on ${\cal B}(\Hi),$ that can be given an operator sum representation (OSR) \cite{kraus}:
\[\Ec(X)=\sum_kM_kXM_k^\dag, \]
where $\dag$ denotes the adjoint operator, or the transpose conjugate when a matrix representation is used. 
A CP map is also TP if and only if $\sum_kM_k^\dag M_k=I.$

Assume that a neighborhood structure is {\em given}, and that each subsystem is contained in some neighborhood.  A CP map $\Ei$ is a \emph{neighborhood map} (with respect to a neighborhood  ${\cal N}_j$) if there exists ${\cal N}_j$ such that 
\beq \Ei=\Ei_{{\cal N}_j}\otimes {\cal I}_{\overline{\cal N}_j},
\eeq 
where $\Ei_{{\cal N}_j}$ is the restriction of $\Ei$ to operators on the subsystems in ${\cal N}_j$ and ${\cal I}_{\overline{\cal N}_j}$
the identity map for operators on $\Hi_{\overline{\cal N}_j}$, respectively.
An equivalent formulation can be given in terms of an OSR: $\Ei(\rho)=\sum_k M_k \rho M_k^\dag$ is a ${\cal N}_j$-neighborhood map 
(or simply QL with respect to ${\cal N}_j$) if there exists a neighborhood
${\cal N}_j$ such that, {\em for all} $k,$ \[M_k = M_{{\cal N}_j,k}\otimes I_{\overline{\cal N}_j}.\] 
The reduced map on the neighborhood then reads \[\Ei_{{\cal N}_j}(\cdot )=\sum_k M_{{\cal N}_j,k}\,\cdot\,M_{{\cal N}_j,k}^\dag.\]
As the identity factor is preserved by sums (and products) of the $M_k$, it is immediate to verify that the QL property is well-defined with
respect to the freedom in the OSR. 

For the discrete-time QL dynamics we are interested in, the relevant stabilizability properties are summarized in the following \cite{Ticozzi-alternating,johnson-FTS}: 

\begin{defin}
A target state $\rho \in {\cal D}(\Hi)$ is \emph{quasi-locally stabilizable} (QLS) with respect to a neighborhood structure $\neigh$ 
if there exists a 
sequence of CPTP neighborhood maps
$\{\mathcal{E}_t\}_{t\geq 0}$ such that:
\begin{eqnarray}
 \mathcal{E}_t (\rho) &\hspace*{-2mm} =\hspace*{-2mm} & \rho, \quad  \forall t\geq 0, 
\label{invFT} \\
\lim_{t\rightarrow +\infty}\mathcal{E}_{t,0}(\sigma)  
& \hspace*{-2mm} \equiv \hspace*{-2mm} &\lim_{t\rightarrow +\infty}(\mathcal{E}_{t}\circ \mathcal{E}_{t-1} \circ \ldots \circ \mathcal{E}_0) (\sigma)\nonumber\\
&\hspace*{-2mm}  =\hspace*{-2mm} & \rho, \quad \forall \sigma \in  \mathcal{D}(\Hi),  
\label{attractQLS}
\end{eqnarray}
A target state $\rho$ is quasi-locally \emph{finite-time stabilizable} (FTS) in $T$ steps if there exists a finite sequence 
$\{\mathcal{E}_t\}_{t=1}^{T}$ of neighborhood maps satisfying the invariance property \eqref{invFT} and ensuring that, 
$\forall \sigma \in  \mathcal{D}(\Hi)$, 
\begin{eqnarray}
\hspace*{-5mm} \mathcal{E}_{T,0}(\sigma)  
&=&(\mathcal{E}_{T}\circ \mathcal{E}_{T-1} \circ \ldots \circ \mathcal{E}_0) (\sigma)= \rho. 
\label{attractFT}
\end{eqnarray}
Furthermore, $\rho$ is \emph{robustly finite-time stabilizable} (RFTS) if \eqref{invFT} and $(\ref{attractFT})$ 
hold for any permutation of the $T$ maps.
\end{defin}

\subsection{Characterization of asymptotic QL stabilizability}
\label{sec:invariance}

We next recall the characterization of QLS pure states given in \cite{Ticozzi-alternating}, which in turns build on 
properly characterizing the interplay between the invariance condition \eqref{invFT} and the QL constraint 
on CPTP dynamics \cite{Ticozzi2012,TicozziQIC2014,Johnson2016}.  This effectively 
impose certain minimal fixed-point set, and hence suggests a structure for the stabilizing dynamics. 

In order to formalize these results, we need to introduce the concept of Schmidt-span. 
Given $X\in \mathcal{B}(\Hi_A\otimes \Hi_B)$, with corresponding (operator) Schmidt decomposition $X=\sum_j A_j\otimes B_j,$
we define the {\em Schmidt span} of $X$ as: $$\Sigma_A(X) \equiv \text{span}( \{ A_j \}).$$

The Schmidt span is important because, if we want to leave an operator invariant with a neighborhood map ${\cal E}$,  this 
also imposes the invariance of all operators with support on its Schmidt span.  Let $\fix({\cal E})$ denote the fixed-point set of ${\cal E}$. 
The following result, proven in \cite{Johnson2016}, makes this idea precise:

\vspace*{1mm}

{\begin{cor} 
Let $\rho \in \mathcal{D} ( \Hi_{ {\cal N}_j} \otimes  \Hi_{\overline{\mathcal N}_j})$ and 
$\mathcal{E}=\mathcal{E}_{{\cal N}_j} \otimes I_{{\overline{\mathcal N}}_j}$ a neighborhood map. Then if 
$$\span(\rho)\subseteq \fix(\mathcal{E}),$$ it must also be that 
$$\Sigma_{\mathcal{N}_j}(\rho)\otimes \mathcal{B}(\Hi_{\overline{\mathcal{N}}_j})\subseteq \fix(\mathcal{E}_{{\cal N}_j}).$$
\end{cor}}

\vspace*{1mm}

In particular, if the target state is pure, $\rho=\ket{\psi}\bra{\psi},$ and the state vector admits a Schmidt decomposition 
of the form $\ket{\psi}=\sum_k \ket{a_j}\otimes\ket{b_j}$ 
with respect to the bipartition $\Hi_{\mathcal{N}_j}\otimes \Hi_{\overline{\mathcal{N}}_j},$ it is possible to show that $\Sigma_{\mathcal{N}_j}(\rho)={\cal B}(\Hi^0_{{\cal N}_j}),$ where $\Hi^0_{{\cal N}_j}  \equiv {\span}(\{\ket{a_j}\})\subseteq \Hi_{\mathcal{N}_j}.$
Leveraging the above observation, the following characterization of QLS pure states has been proved in \cite{Ticozzi-alternating}:

\vspace*{1mm}

\begin{thm}[QLS pure states]
\label{mainthm} 
A pure state $\rho=|\psi\rangle\langle \psi|$ is QLS by discrete-time dynamics if and only if 
\begin{equation}
\label{GAScond}
\supp(\rho) = \bigcap_j {\mathcal H}^0_j  \equiv \bigcap_j ( \Hi^0_{{\cal N}_j} \otimes \Hi_{\overline\Ni_j}) .
\end{equation}
\end{thm}

\subsection{QL stabilization as frustration-free cooling}
\label{sec:physint}
 
In order to develop some physical intuition on the role of $\Hi_j^0$ in the Theorem above, and understand how the maps used 
in the proof attain stabilization, it is convenient to resort to the concept of a parent Hamiltonian. 

Consider a QL Hamiltonian, namely, $H=\sum_k H_k,$ with 
$H_k=H_{{\cal N}_k}\otimes I_{\overline{\cal N}_k}.$ 
{$H$ is called a {\em parent} Hamiltonian for a pure state $\ket{\psi}$ 
if it admits $\ket{\psi}$ as a ground state, and it is 
called a {\em frustration-free (FF)} Hamiltonian if any global 
ground state is {\em also} a local ground state, that is, 
\[ \text{argmin}_{\ket{\psi} \in \Hi} \bra{\psi} H \ket{\psi} \subseteq 
\text{argmin}_{\ket{\psi} \in \Hi} \bra{\psi} H_k \ket{\psi}, \forall k. \] }

\noindent Suppose that a target state $\ket{\psi}$ admits a FF QL parent Hamiltonian $H$ for which it is 
the {\em unique} ground state. 
Then, similarly to what is possible for continuous-time dissipative preparation by Markovian 
dynamics \cite{kraus-dissipative,Ticozzi2012}, the structure of $H$ may be naturally used to derive a
stabilizing discrete-time dynamics: it suffices to implement neighborhood maps that locally ``cool'' the system. 
This is done in the proof of Theorem \ref{mainthm}. 

The following corollary follows \cite{Ticozzi-alternating}:

\begin{cor}
\label{parent}
{A state $\rho=|\psi\rangle\langle\psi|$ is QLS by discrete-time dynamics if and only if it is 
the unique ground state of a FF QL parent Hamiltonian.}
\end{cor}

\vspace*{1mm}

Among the possible QL FF parent Hamiltonians that a given pure state may admit, 
one can be constructed in a {\em canonical way} from the state itself as follows:
\begin{defin}
\label{def:canham}
Given a neighborhood structure $\neigh =\{ {\cal N}_j \}$, the \emph{canonical FF parent Hamiltonian} 
associated to $|\psi\rangle$ is 
\begin{equation}
H_{\ket{\psi}} \equiv \sum_{j} (I - \Pi_{\neigh_j}\otimes I_{\overline{\neigh}_j}) \equiv \sum_j (I-\Pi_j),
\label{canonicalH}
\end{equation}
in terms of the 
projectors $\Pi_{\neigh_j}$ and $\Pi_j$ onto the Schmidt span 
$\Hi^0_{\neigh_j}$ and the extended Schmidt span $\overline\Hi^0_j$, respectively. 
\end{defin}

\vspace*{1mm}

The canonical parent Hamiltonian plays an important role in the characterization of RFTS states for the locality 
of interest, in particular, in reference to its commutativity properties: we say that $H_{\ket{\psi}}$ is {\em commuting} if 
the defining projectors commute, $[\Pi_k,\Pi_j]=0$ for all $j,k.$

\subsection{Sufficient conditions for RFTS} 

While QLS pure states admit a characterization that is both intuitive and lends itself to design of stabilizing dynamics, an equivalent 
result is not available for FTS or RFTS at a similar level of generality. In \cite{johnson-FTS}, a number of sufficient conditions are provided. 
Here, we focus on the most general one that ensures RFTS, and that is also the most interesting towards unsupervised and distributed control  implementation.

The key idea behind the relevant RFTS sufficient condition is to identify a decomposition of the full Hilbert space into {\em virtual subsystems}, such that (1) the QL constraint is respected; (2) the target state looks like a {\em virtual product state}. In order to obtain such a decomposition 
of $\Hi$, two steps may be required in general:

\vspace{2mm}
\noindent $\bullet$ {\bf Coarse graining:}
First, we group physical subsystems into coarse-grained ones, that are contained in the same neighborhoods. Formally:

\begin{defin}
Given $\Hi\cong\bigotimes_{i=1}^n \Hi_i$ and a neighborhood structure $\neigh$, we define \emph{coarse-grained subsystems} (or 
coarse-grained particles) $v$ to be the subsets of physical subsystems $i$ such that $i,j\in v$ when $i\in \Ni_k$ implies $j\in\Ni_k$; that is, $v$ are the group of subsystems that are contained in exactly the same set of neighborhoods. We define the coarse-grained subsystem 
space as $\Hi_v\equiv \bigotimes_{i\in v}\Hi_i.$ 
\end{defin}

\vspace*{1mm}

\noindent
Coarse-graining of physical subsystems allows us to aggregate groups of subsystems that are subject to exactly the same constraints as far as the neighborhood structure is concerned, and on which we have full control. It is easy to see that, by construction, coarse-grained subsystems are mutually disjoint, in the sense that no physical subsystem can belong to two different coarse-grained subsystems.

\vspace{2mm}

\noindent $\bullet$ {\bf Local reduction:}
Next, we reduce each coarse-grained subsystem space to a subspace. Formally:

\begin{defin}
Consider $\Hi\cong \bigotimes_{v}\Hi_v,$ and identify a set of subspaces $\Hi^{\ell}_v\leq\Hi_v$. The {\em locally restricted space with respect to the $\Hi_v^\ell$} is then \(\Hi^{\ell}\cong\bigotimes_{v}\Hi^{\ell}_v.\)
In particular, if $\rho\in{\frak D}(\Hi)$ and $\Hi^{\ell}_v \equiv \supp(\text{Tr}_{\overline v}(\rho))\leq\Hi_v$, the $\Hi^{\ell}$ is the
{\em locally restricted space with respect to $\rho$.}
\end{defin}

\vspace*{1mm}

\noindent 
We are now ready to quote a set of sufficient conditions for a pure state to be RFTS, that covers all 
the states we know, or we can construct \cite{johnson-FTS}:

\begin{thm}
{\em (Neighborhood factorization on local restrictions):}
\label{thm:lrhilbertdecomp}
A pure state $\ket{\psi}$  is RFTS relative to $\neigh$ if there exist a locally restricted space $\Hi^{\ell}\cong\bigotimes_{i} \Hi^{\ell}_i$,  
containing the support of the target state, and a unitary change of basis 
$U:\Hi \rightarrow \Hi^0\oplus (\bigotimes_{j=1}^M \hat{\Hi}_j)$ such that:
\begin{enumerate} 
\item The local restriction is respected, that is, $(\Hi\ominus\Hi^{\ell})$ is invariant for $U$; 
\item 
The target state becomes a virtual product state:
\begin{equation}
U\ket{\psi}=0\oplus \Big(\bigotimes_{j=1}^M\ket{\hat{\psi}_j} \Big);
\end{equation}
\item The QL constraint is respected, that is, for each virtual subsystem factor $\hat{\Hi}_j$, there exists a neighborhood $\neigh_k$ 
such that
\begin{equation}
U^{-1}(I^0\oplus\mathcal{B}(\hat{\Hi}_j)\otimes I_{\overline{j}})\,U \leq \mathcal{B}(\Hi_{\neigh_k})\otimes I_{\overline{\neigh}_k}.
\end{equation}
\end{enumerate}
\end{thm}

\vspace{2mm}
\noindent 

A number of remarks are in order.

First, given condition 2) above, $\ket{\psi}$ is the unique ground state of a Hamiltonian 
\[ \hat H=\sum_k \hat H_k, \;\;
\hat H_k=0\oplus\Big( \ket{\hat{\psi}_j}\bra{\hat{\psi}_j} \otimes\bigotimes_{\ell\neq j}^M I_j\Big),\] 
where each $\hat H_k$
projects onto the complement of the span of $\ket{\hat\psi_j}.$ Given the ``virtually factorized'' structure of the state, 
these projections {\em commute}. Furthermore, given property 3), these are all neighborhood Hamiltonians. Hence, $\hat H$ is a parent Hamiltonian for $\ket{\psi}$ with respect to the given neighborhood structure, and is, in fact, its {canonical} parent Hamiltonian. 
 
Second, if the above conditions 1)--3) hold, it is easy to see that a CPTP map that prepares $ \ket{\hat{\psi}_j}\bra{\hat{\psi}_j}$ 
not only exists but it is a neighborhood map. This directly gives RFTS. 
However, it is in general hard to {\em check} if a state admits such a decomposition, or, equivalently, to find the transformation 
$U$ in Theorem \ref{thm:lrhilbertdecomp}.

\section{Characterization of  robust finite-time stabilizability for tree-like hypergraphs}

While Theorem \ref{thm:lrhilbertdecomp} covers all the known examples of RFTS states (see \cite{johnson-FTS} for a more in-depth discussion), as yet we have no indication on how to find a good $U$ for a given $\rho$, nor do we know whether the above sufficient conditions are also necessary. We now describe a class of QL constraints for which a neighborhood factorization is {\em both} sufficient and necessary, as well as equivalent to the existence of a {\em commuting} canonical parent Hamiltonian. 

\subsection{Tree-like hypergraphs}\label{ssec:treelike}
The class of QL constraints we focus on satisfies two key properties.
The first constrains the way in which neighborhoods can overlap: 

\begin{defin}
A neighborhood structure satisfies the \emph{matching overlap (MO)} condition if for any set of neighborhoods 
that have a common intersection, this common intersection is also the intersection of any pair of the 
neighborhoods in the set.
\end{defin}

\vspace*{1mm}

\begin{figure}[t]
\begin{center}
\includegraphics[width=0.5\columnwidth]
{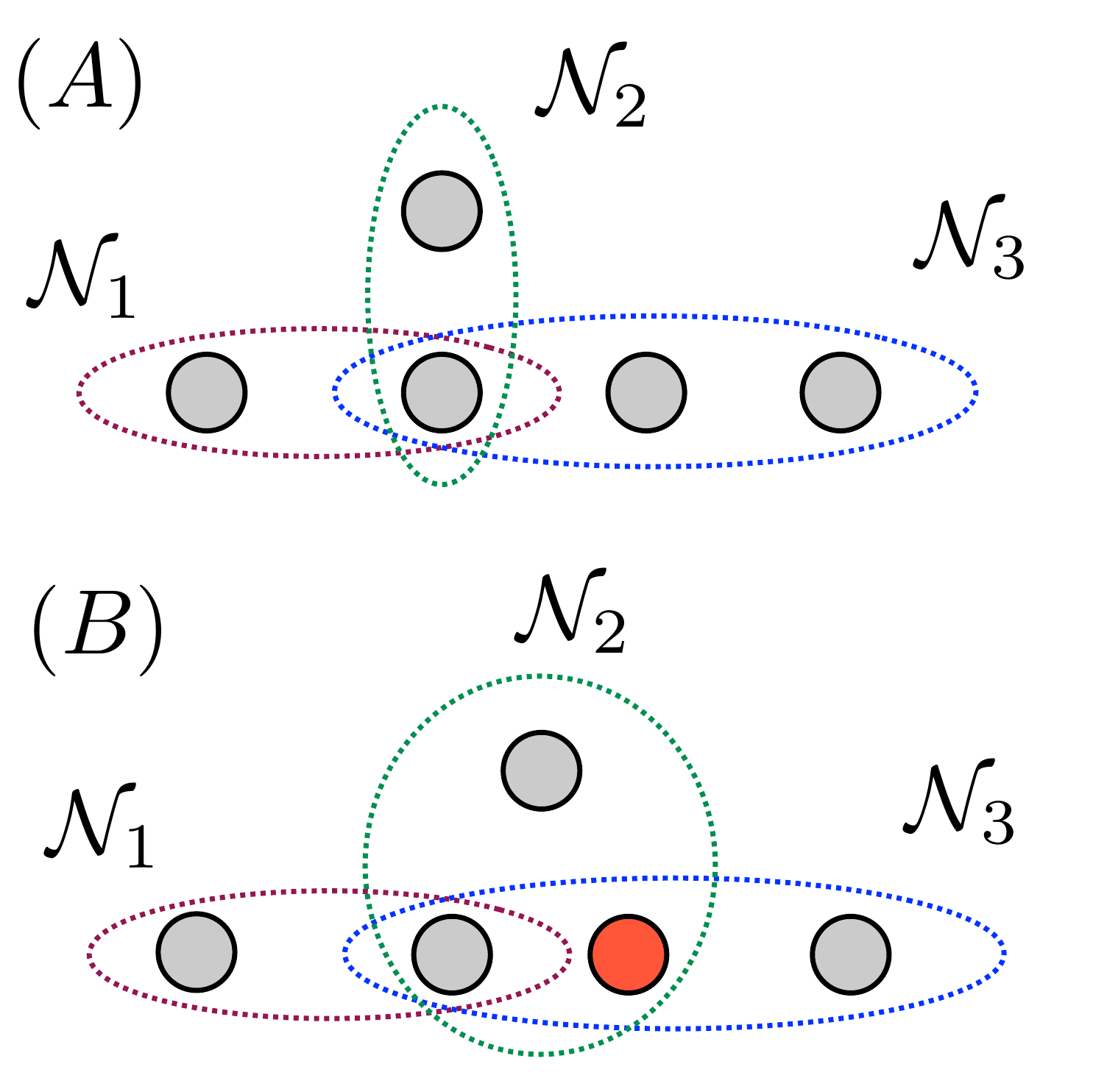}
\end{center}
\vspace*{-3mm}
\caption{(A) shows a neighborhood structure that obeys the MO property, as the intersection of any pair of $\Ni_j$ is also 
the intersection of the triple. By enlarging ${\cal N}_2$ to include the red subsystem, (B) fails to 
satisfy MO instead.} 
\label{fig:MOvsNOMO}
\end{figure}

This property, in turn, implies that the neighborhoods can only intersect on a single coarse-grained particle. While two-body neighborhoods necessarily satisfy the MO condition, general neighborhood structures need not. Figure \ref{fig:MOvsNOMO} further illustrates the MO property: 
in both panels (A) and (B) the three neighborhoods have a non-empty intersection; however, in (B), the intersection of the pair $\Ni_2$ and 
$\Ni_3$ contains an extra subsystem, which causes the MO to fail.

The second property we need to impose is the absence of ``cycles''. In order to formalize it, we need to define what is a path on 
the multipartite system that is compatible with $\neigh$, or, equivalently a path on the associated hypergraph.
A (finite) {\em path} on $\neigh$ is a finite sequence of subsystem indexes interspaced by neighborhoods: 
\[j(0),\Ni_{k(1)},j(1),\Ni_{k(2)},j(2),\ldots,\Ni_{k(N)},j(N),\] that satisfies:
\begin{itemize}
\item $j(\ell)\neq j(m)$ for all $(\ell,m)\neq(0,m);$
\item $\Ni_{k(\ell)}\neq\Ni_{k(m)}$ for all $(\ell,m);$
\item $j(\ell-1),j(\ell)\in \Ni_{k(\ell)}$ for all $\ell=1,\ldots,N.$
\end{itemize}
We the say that subsystem $a$ is connected to subsystem $b$ if there is a path with $j(0)=a, j(M)=b,$ and that a neighborhood 
{\em $\Ni_j$ is connected to $\Ni_\ell$} if there exists a path from $a\in\Ni_j$ to some $b\in\Ni_\ell.$ If $j(0)= j(M)$ it is called a 
{\em cycle path}. We then have the following:

\begin{defin}
A neighborhood structure (an hypergraph) is {\em tree-like} if it obeys the MO property and is acyclic, 
that is, it does not allow for cycle paths.
\end{defin}

\vspace*{1mm}

Notice that, if the MO property holds, each neighborhood contains coarse-grained particles that belong either to that neighborhood alone, or to an intersection. One can then construct a (standard) {graph} by removing the particles that belong to a single neighborhood, associating nodes to the multi-neighborhood particles and adding edges between pairs of the latter that belong to the same neighborhood. If the resulting graph admits cycle paths (in the standard sense) then the initial hypergraph does as well, and hence it is not tree-like, and viceversa.

\begin{figure}[t]
\hspace{6mm}
\includegraphics[width=0.8\columnwidth]
{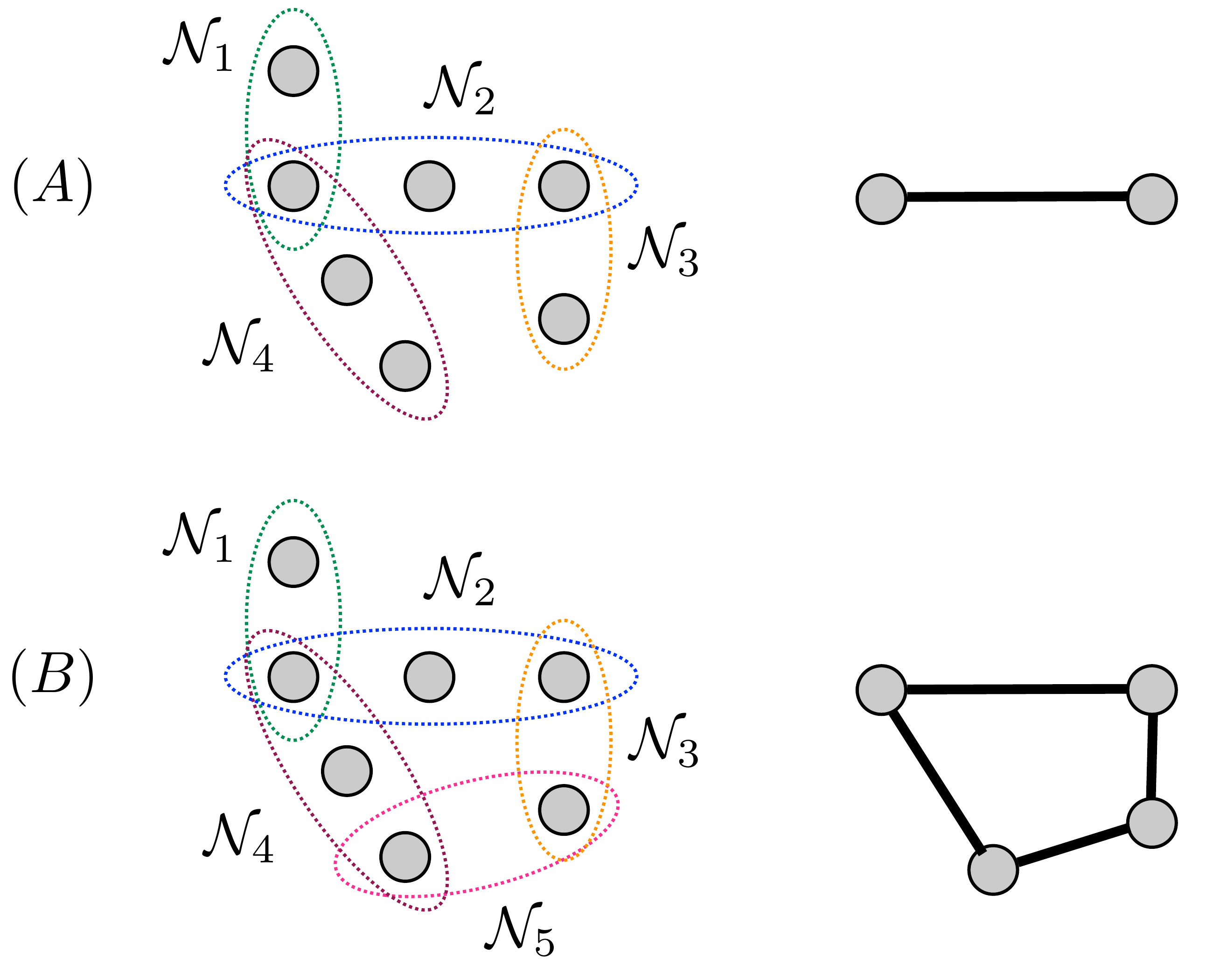}
\vspace*{-3mm}
\caption{(A) shows a MO neighborhood structure that is acyclic; by adding neighborhood ${\cal N}_5$, (B) gains a cycle. 
On the right side, we depict the corresponding graphs after the single-neighborhood subsystems are removed, which can be equivalently used to determine the tree-like property.} 
\label{fig:noloops}
\end{figure}

\subsection{Main result}

The main result of this paper can now be stated as follows:

\begin{thm}[RFTS on trees]
\label{thm:treenecessity}
Let $\neigh$ be a tree-like neighborhood structure on $\Hi$. A pure state $\ket{\psi}$ is RFTS with respect to $\neigh$ if and only 
if the projectors $\Pi_j$ onto the neighborhood reduced states of $\ket{\psi}$ commute pairwise.
\end{thm}

\vspace*{1mm}

The ``if'' implication follows directly from Theorem V.13 of \cite{johnson-FTS}. We thus focus on the necessity part. 
The proof requires a few auxiliary results which have also been proved in \cite{johnson-FTS}, and which we recall next.
The first result constrains the form of a stabilizing neighborhood map:

\begin{lemma}
\label{thm:invarianceoutputlem}
If $\mathcal{E}_k$ acting on $\neigh_k$ preserves $\ket{\psi}\bra{\psi}$, then,  
for arbitrary $\rho$ it must be 
\begin{equation}
\Pi_k \mathcal{E}_k(\rho) \Pi_k = \Pi_k \rho \Pi_k + \Ei_R(\Pi_k^{\perp}\rho\Pi_k^{\perp}),
\end{equation}
for some CPTP $\Ei_{R}.$
\end{lemma}

\vspace*{1mm}

\noindent 
We will also make use of the following trace inequality:
\begin{prop}
Let $\Pi_1$ and $\Pi_2$ be projectors, with $\Pi_{1\cap 2}$ the projector onto their intersection. Then
\begin{equation}
\Tr (\Pi_1\Pi_2) \geq \Tr(\Pi_{1\cap 2})+\frac{1}{2}\Tr(|[\Pi_1,\Pi_2]|^2).
\end{equation}
\end{prop}

\vspace*{1mm}

\noindent 
Combining the two results above, the following result concerning the commutativity of the projections onto the supports 
is established in  \cite{johnson-FTS}. 
Define ${\Ni}_{\widecheck{k}} \equiv \bigcup_{j\neq k} \neigh_j.$
\begin{prop}
\label{thm:complementcommutecondition}
If $\ket{\psi}$ is RFTS with respect to neighborhood structure $\neigh$, then $[\Pi_k,\Pi_{\widecheck{k}}]=0$ for all neighborhoods $\neigh_k$, where $\Pi_k$ and $\Pi_{\widecheck{k}}$ are the projectors
onto 
$\Hi_k^0=\textup{supp}(\Tr_{\overline{k}}(\ket{\psi}\bra{\psi})\otimes\identity_{\overline{k}})$ and $\Hi_{\widecheck{k}}^0=\textup{supp}(\Tr_{j\notin \Ni_{\widecheck{k}}}(\ket{\psi}\bra{\psi})\otimes\identity_{j\notin \Ni_{\widecheck{k}}}),$ respectively.
\end{prop}

\vspace*{1mm}

Here, we prove an alternative necessity result showing that RFTS does require commutativity of projectors {\em beyond} just $\Pi_k$ 
and $\Pi_{\widecheck{k}}$ as above; in fact, commutativity is required essentially for any pair of projections emerging from a bipartition of the neighborhoods: 

\begin{prop}
\label{thm:complementcommuteconditionold}
Let $\ket{\psi}$ be RFTS with respect to $\neigh,$ $\Ni_\Lambda \equiv \bigcup_{j\in\Lambda}\neigh_j$ be the union of neighborhoods with index in some subset $\Lambda,$ and ${\Ni}_{\widecheck\Lambda} \equiv \bigcup_{j\notin \Lambda} \neigh_j.$ Then $[\Pi_\Lambda,\Pi_{\widecheck{\Lambda}}]=0,$ where $\Pi_\Lambda$ and $\Pi_{\widecheck{\Lambda}}$ orthogonally project onto $\textup{supp}(\Tr_{j\notin \Lambda}(\ket{\psi}\bra{\psi})\otimes\identity_{j\notin{\Lambda}})$ and $\textup{supp}(\Tr_{j\notin \widecheck{\Lambda}}(\ket{\psi}\bra{\psi})\otimes\identity_{j\notin \widecheck{\Lambda}})$, respectively.
\end{prop}

\vspace*{1mm}

\proof
Assume that $\ket{\psi}$ is RFTS by neighborhood maps $\mathcal{E}_k$. Without loss of generality, assume that the subset of maps indexed by $\Lambda$ acts after the remaining neighborhood maps. Let $\mathcal{E}_\Lambda$ be the composition of this subset of maps and $\mathcal{E}_{\widecheck{\Lambda}}$ be the composition of the remaining neighborhood maps. Robust stabilizability, then, implies that 
$\mathcal{E}_{\Lambda}\circ\mathcal{E}_{\widecheck{\Lambda}}(\cdot)=\ket{\psi}\bra{\psi}\Tr(\cdot)$. By the invariance requirement, $\mathcal{E}_{\widecheck{\Lambda}}(\Pi_{\widecheck{\Lambda}})=\Pi_{\widecheck{\Lambda}}$. Thus, applying the sequence to $\Pi_{\widecheck{\Lambda}}$, we have
\begin{eqnarray*}
\ket{\psi}\bra{\psi}\Tr(\Pi_{\widecheck{\Lambda}})=\mathcal{E}_{l}\circ\mathcal{E}_{\widecheck{\Lambda}}(\Pi_{\widecheck{\Lambda}}) 
= \mathcal{E}_{l}(\Pi_{\widecheck{\Lambda}}).
\end{eqnarray*}
Conjugating both sides of the equation with respect to $\Pi_\Lambda$, we can apply Lemma \ref{thm:invarianceoutputlem} to obtain
\begin{eqnarray*}
\ket{\psi}\bra{\psi}\Tr(\Pi_{\widecheck{\Lambda}})=\Pi_\Lambda\Pi_{\widecheck{\Lambda}}\Pi_\Lambda+\Pi_\Lambda\sigma\Pi_\Lambda,
\end{eqnarray*}
where $\sigma$ is some positive-semidefinite operator. Next, conjugating both sides of the new equation with respect to the projector $\tilde{\Pi}_\Lambda\equiv\Pi_\Lambda-\ket{\psi}\bra{\psi}$ makes the left-hand side equal to zero, while leaving the sum of two positive semidefinite operators on the right hand side, namely,
\begin{eqnarray*}
0=\tilde{\Pi}_\Lambda\Pi_\Lambda\Pi_{\widecheck{\Lambda}}\Pi_\Lambda\tilde{\Pi}_\Lambda+\tilde{\Pi}_\Lambda\Pi_\Lambda\sigma\Pi_\Lambda\tilde{\Pi}_\Lambda.
\end{eqnarray*}
The sum of two positive-semidefinite matrices is zero only if both matrices are zero. Taking the trace of the first zero matrix thus gives
\begin{eqnarray*}
0&=&\Tr(\tilde{\Pi}_\Lambda\Pi_\Lambda\Pi_{\widecheck{\Lambda}})\\
&=&\Tr((\Pi_\Lambda-\ket{\psi}\bra{\psi})\Pi_{\widecheck{\Lambda}})\\
&=&\Tr(\Pi_\Lambda\Pi_{\widecheck{\Lambda}})-\Tr(\Pi_{\Lambda\cap{\widecheck{\Lambda}}})\\
&\geq&\Tr(|[\Pi_\Lambda,\Pi_{\widecheck{\Lambda}}]|^2).
\end{eqnarray*}
This holds only if $[\Pi_\Lambda,\Pi_{\widecheck{\Lambda}}]=0$. As the above arguments are made for a general index subset $\Lambda$, 
they must hold for all such index sets. \qed

\noindent 

We next prove a proposition that allows for a simplification of the commutation condition in Proposition \ref{thm:complementcommutecondition}
for certain neighborhood structures: 

\begin{prop}
\label{thm:commprojreduce1}
Let $A$, $B$, and $C$ be three neighborhoods, such that $A\cap B\subseteq C$. 
Let $\Pi_A$ and $\Pi_B$ be two neighborhood projectors on $A$ and $B$, respectively, and define, in addition, the projectors $\Pi_{\Tr_{\overline{C}}(\Pi_A)}$ on $\textup{supp}(\Tr_{\overline{C}}(\Pi_A)\otimes \identity_{\overline{C}})$ and $\Pi_{\Tr_{\overline{C}}(\Pi_B)}$ on $\textup{supp}(\Tr_{\overline{C}}(\Pi_B)\otimes \identity_{\overline{C}})$. Then: 
\beq[\Pi_A,\Pi_B]=0\implies[\Pi_{\Tr_{\overline{C}}(\Pi_A)},\Pi_{\Tr_{\overline{C}}(\Pi_B)}]=0.\eeq
\end{prop}

\vspace*{2mm}
\proof 
 From $[\Pi_A,\Pi_B]=0$, we have $\Tr_{\overline{C}}([\Pi_A,\Pi_B])=0$. Let $\sigma_{\Tr_{\overline{C}}(\Pi_A)}\equiv \Tr_{\overline{C}}(\Pi_A)$ and $\sigma_{\Tr_{\overline{C}}(\Pi_B)}\equiv \Tr_{\overline{C}}(\Pi_B)$. Since $C$ contains $A\cap B$, we have $[\sigma_{\Tr_{\overline{C}}(\Pi_A)},\sigma_{\Tr_{\overline{C}}(\Pi_B)}]=0$. Finally, since $\Pi_{\Tr_{\overline{C}}(\Pi_A)}$ can be written using the same resolution of the identity of $\sigma_{\Tr_{\overline{C}}(\Pi_A)}$, and similarly for $\Pi_{\Tr_{\overline{C}}(\Pi_B)}$ with respect to $\sigma_{\Tr_{\overline{C}}(\Pi_B)}$, these must also commute: $[\Pi_{\Tr_{\overline{C}}(\Pi_A)},\Pi_{\Tr_{\overline{C}}(\Pi_B)}]=0$.
\qed

In our application of the above result, both $\Pi_A$ and $\Pi_B$ will be supports of reduced states, say, $\rho_A$ and $\rho_B$. 
However, the objects we are ultimately concerned with are of the form $\Pi_{\Tr_{\overline{C}}(\rho_A)}$, 
which is not, {\em a priori}, the same as the $\Pi_{\Tr_{\overline{C}}(\Pi_A)}$ of the above proposition. 
In order to apply the above proposition to projectors on neighborhood-reduced states, we prove that, in fact, 
$\Pi_{\Tr_{\overline{C}}(\rho_A)}=\Pi_{\Tr_{\overline{C}}(\Pi_A)}$. The following proposition suffices:

\begin{prop}
\label{thm:convexreducedsupport}
Let $\rho \in\mathfrak{D}(\Hi_1\otimes\Hi_2)$ be a positive semidefinite operator,  
with spectral decomposition $\rho=\sum_i \lambda_i P_i$.  Then $\textup{supp}(\Tr_2(\rho))=\textup{supp}(\Tr_2(\sum_{ij}r_j\lambda^{\alpha_j}P_i))$, for $r_j>0$ and $\alpha_j\in \mathbb{R}.$ 
\end{prop}

\proof
Consider the spectral decomposition $\rho=\sum_i \lambda_i P_i$ with $\lambda_i \geq 0$. We then compute
\begin{eqnarray*}
\textup{supp}(\Tr_2(\rho))
&=& \textup{supp}(\sum_i \lambda_i\Tr_2(P_i))\\
&=&\bigcup_i\textup{supp}(\lambda_i\Tr_2(P_i))\\
&=&\bigcup_i\textup{supp}(\sum_jr_j\lambda_i^{\alpha_j}\Tr_2(P_i))\\
&=&\textup{supp}(\Tr_2(\sum_{ij} r_j \lambda_i^{\alpha_j}P_i)),
\end{eqnarray*}
for any $\alpha_j\in\mathbb{R},$ and $r_j> 0$. \qed

\vspace*{1mm}
\noindent 
In particular, setting $r_j=\delta_{ij}$, with $\alpha_i=0$, we have $\sum_{ij} r_j \lambda_i^{\alpha_j}P_i=\sum_i P_i = 
\Pi$, the orthogonal projection onto $\supp(\rho)$. Hence, $\textup{supp}(\Tr_2(\rho)) = \textup{supp}(\Tr_2(\Pi))$.

\vspace{2mm}\begin{cor}
\label{thm:commprojpairreduce2}
Let $\ket{\psi}$ be a many-body pure state. Let $A$ and $B$ be two, possibly overlapping, neighborhoods. Let $C$ be a neighborhood containing $A\cap B$. For any neighborhood $X$, let $\Pi_X$ be the projector onto $\textup{supp}(\Tr_{\overline{X}}(\ket{\psi}\bra{\psi})\otimes\identity_{\overline{X}})$. If $[\Pi_A,\Pi_B]=0$, then $[\Pi_{A\cap C},\Pi_{B\cap C}]=0$.
\end{cor}
\proof
While Proposition \ref{thm:commprojreduce1} ensures the commutation of the projection onto the support of  $\Tr_{\overline C}(\Pi_A),$ and similarly for $B$, here $\Pi_{A\cap C}$ denotes the projection onto the support of $\ketbra{\psi}$ in $A \cap C$. However, the particular use of Proposition \ref{thm:convexreducedsupport} outlined above with $r_j=\delta_{ij}$ and $\alpha_i=0$ ensures that $\textup{supp}(\Tr_{\overline{C}}(\Tr_{\overline{A}}(\ketbra{\psi})) = \textup{supp}(\Tr_{\overline{C}}(\Pi_{A})),$ proving that the two projections are the same.
\qed

Finally, by exploitng the MO and the acyclic properties, as well as Corollary \ref{thm:commprojpairreduce2}, we can refine 
the necessary condition of  Proposition \ref{thm:complementcommutecondition} for RFTS, as anticipated:

\vspace{2mm}

\noindent {\em Proof of the ``only if'' implication of Theorem \ref{thm:treenecessity}}.
Assuming that $\ket{\psi}$ is RFTS, Proposition \ref{thm:complementcommutecondition} implies that $[\Pi_\Lambda,\Pi_{\widecheck{\Lambda}}]=0$ for any bipartition of the set of neighborhoods in $\Lambda,\overline{\Lambda}$: we here use the hypothesis on the QL notion in order to show that pairwise commutativity of the $\Pi_j$ is actually necessary as well. Consider two neighborhoods $\neigh_j$ and $\neigh_k$. If $\neigh_j\cap\neigh_k=\{\}$, then $[\Pi_j,\Pi_k]=0$ trivially. In the case that they do intersect, define $a\equiv\neigh_j\cap\neigh_k$. Thanks to MO property,  $a$ is a single coarse-grained particle. Define $A$ to be the union of neighborhoods which are connected to $\neigh_j$ by a path that starts with $a,\neigh_j,\ldots.$ The tree-like property then guarantees that $A$ cannot contain any other neighborhood that contains $a$; hence, $A$ will not be the full neighborhood set: in fact, assume by contradiction that $A$ contains some $\neigh_\ell,$ such that $a\in\neigh_\ell$; by definition of $A$ there would exist a cycle starting with $a,\neigh_j,\ldots$ that ends with $\ldots,\neigh_\ell, a,$ which is impossible by hypothesis. Hence, the disjoint sets of neighborhoods $B=\overline A$ and $A$ are non empty and form a bipartition of the set of neighborhoods. By identifying $A=\Lambda,B=\widecheck{\Lambda},$ Proposition \ref{thm:complementcommutecondition} ensures that  $[\Pi_A,\Pi_{B}]=0$. Moreover, $B$ contains $\neigh_k$. Let $C\equiv \neigh_j\cup \neigh_k,$ so $C\supseteq \{a\}=A\cap B$. Hence, we may apply Corollary \ref{thm:commprojpairreduce2} to obtain $[\Pi_j,\Pi_k]=0$ from $[\Pi_A,\Pi_{B}]=0$.
\qed


Now notice that the canonical Hamiltonian $H_{\ket{\psi}}$ is composed of QL terms of the form $I-\Pi_k.$ We thus immediately 
obtain the following corollary:

\vspace{1mm}
\begin{cor}\label{cor:commham} Let $\neigh$ be a tree-like neighborhood structure. A pure state $\ket{\psi}$ is 
RFTS with respect to $\neigh$ if and only if it is the unique ground state of its canonical FF commuting parent Hamiltonian. 
\end{cor}

\subsection{An example and a further characterization: GBV states}

\begin{figure}[t]
\hspace{-3mm}
\begin{center}
\includegraphics[width=0.85\columnwidth]
{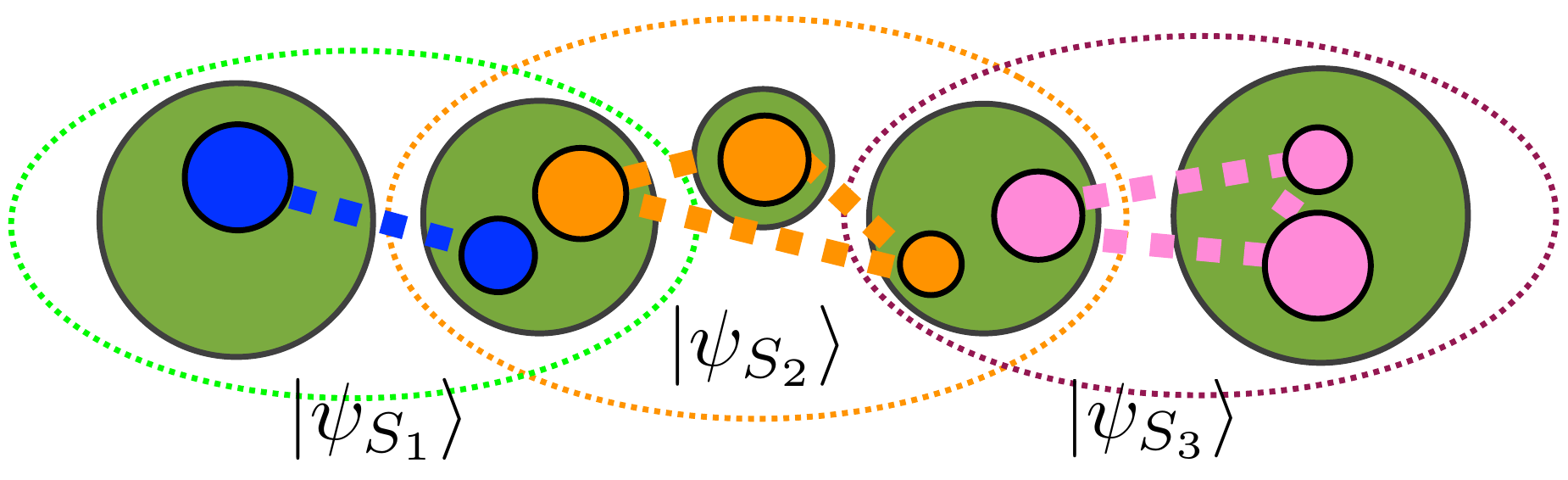}
\end{center}
\vspace*{-3mm}
\caption{Construction of GBV states: virtual particles (colors) are identified within coarse grained ones (green), grouped into $S_k$ (virtual particles of the same color) and a pure entangled state $|\psi_{S_k}\rangle$ is associated to each group.} 
\label{fig:gbv}
\end{figure}

A class of states introduced in \cite{johnson-FTS}, inspired by the work of Bravyi and Vyalyi in \cite{Bravyi2005} and named accordingly, is that of {\em generalized Bravyi-Vyalyi} (GBV) states.
These are constructed by decomposing each coarse-grained particle $v$  into a set of virtual particles, $\Hi_v \equiv \Hi^0_v\oplus \bigotimes_{j=1}^{f_v} \Hi_{vj}$. Then for each neighborhood $\neigh_k$, a subset $S_k$ of pairs $vj$ belonging to that neighborhood is selected such that the sets $S_k$ are disjoint and each $vj$ is contained in some $S_k$. The GBV state is constructed by assigning a pure state factor to each group of $S_k,$ namely $\ket{\psi_{GBV}}=\bigotimes_k\ket{\psi_{S_k}}.$ If the $\ket{\psi_{S_k}}$ are entangled, and the virtual particles do not factorize with respect to the physical subsystem decomposition, $\ket{\psi_{GBV}}$ will be generically non-trivially entangled. 

This method is illustrated for a tree-like hypergraph in Fig. \ref{fig:gbv}. It is easy to see that the canonical parent Hamiltonian for such states 
has QL terms of the form $H_k=I-\ketbra{\psi_{S_k}},$ which commute since they belong to disjoint groups of virtual particles. Hence, on the one hand Corollary \ref{cor:commham} confirms that these states are RFTS.
On the other hand, by extending the $C^*$-algebraic construction in the proof of Theorem V.13 in \cite{johnson-FTS} to leverage the absence of 
cycles in tree-like hypergraphs, it is possible to show that having commuting parent Hamiltonians within this class of QL notions always ensures the existence of a virtual-particle decomposition as the one described above. In other words, a state is RFTS on a tree-like hypergraph 
(equivalently, a state is the unique ground state of a commuting FF QL parent Hamiltonian) if and only if 
it admits a representation as a GBV state.

\bibliographystyle{IEEEtran}
\bibliography{robustbib-1}

\begin{thebibliography}{10}
\providecommand{\url}[1]{#1}
\csname url@samestyle\endcsname
\providecommand{\newblock}{\relax}
\providecommand{\bibinfo}[2]{#2}
\providecommand{\BIBentrySTDinterwordspacing}{\spaceskip=0pt\relax}
\providecommand{\BIBentryALTinterwordstretchfactor}{4}
\providecommand{\BIBentryALTinterwordspacing}{\spaceskip=\fontdimen2\font plus
\BIBentryALTinterwordstretchfactor\fontdimen3\font minus
  \fontdimen4\font\relax}
\providecommand{\BIBforeignlanguage}[2]{{%
\expandafter\ifx\csname l@#1\endcsname\relax
\typeout{** WARNING: IEEEtran.bst: No hyphenation pattern has been}%
\typeout{** loaded for the language `#1'. Using the pattern for}%
\typeout{** the default language instead.}%
\else
\language=\csname l@#1\endcsname
\fi
#2}}
\providecommand{\BIBdecl}{\relax}
\BIBdecl

\bibitem{Viola2001}
S.~Lloyd and L.~Viola, ``Engineering quantum dynamics,'' \emph{Phys. Rev. A},
  vol.~65, p. 010101, 2001.

\bibitem{Cirac2009}
F.~Verstraete, M.~M. Wolf, and J.~I. Cirac, ``Quantum computation and
  quantum-state engineering driven by dissipation,'' \emph{Nature Phys.},
  vol.~5, p. 633, 2009.

\bibitem{schon2005}
C.~Sch\"on, E.~Solano, F.~Verstraete, J.~I. Cirac, and M.~M. Wolf, ``Sequential
  generation of entangled multiqubit states,'' \emph{Phys. Rev. Lett.},
  vol.~95, p. 110503, 2005.

\bibitem{kraus-dissipative}
B.~Kraus, H.~P. B\"uchler, S.~Diehl, A.~Kantian, A.~Micheli, and P.~Zoller,
  ``Preparation of entangled states by quantum markov processes,'' \emph{Phys.
  Rev. A}, vol.~78, p. 042307, 2008.

\bibitem{Cirac2011}
H.~Krauter, C.~A. Muschik, K.~Jensen, W.~Wasilewski, J.~M. Petersen, J.~I.
  Cirac, and E.~S. Polzik, ``Entanglement generated by dissipation and steady
  state entanglement of two macroscopic objects,'' \emph{Phys. Rev. Lett.},
  vol. 107, p. 080503, 2011.

\bibitem{Barreiro-Nature:11}
J.~T. Barreiro, M.~M{\"u}ller, P.~Schindler, D.~Nigg, T.~Monz, M.~Chwalla,
  M.~Hennrich, C.~F. Roos, P.~Zoller, and R.~Blatt, ``An open-system quantum
  simulator with trapped ions,'' \emph{Nature}, vol. 470, no. 7335, pp.
  486--491, 2011.

\bibitem{Yin2013}
Y.~Lin, J.~P. Gaebler, F.~Reiter, T.~R. Tan, R.~Bowler, A.~S. Sorensen,
  D.~Leibfried, and D.~J. Wineland, ``Dissipative production of a maximally
  entangled steady state of two quantum bits,'' \emph{Nature}, vol. 504, p.
  415, 2013.

\bibitem{Devoret}
S.~Shankar, M.~Hatridge, Z.~Leghtas, K.~M. Sliwa, A.~Narla, U.~Vool, S.~M.
  Girvin, L.~Frunzio, M.~Mirrahimi, and M.~H. Devoret, ``Stabilizing
  entanglement autonomously between two superconducting qubits,''
  \emph{Nature}, vol. 504, p. 419, 2013.

\bibitem{Siddiqi}
M.~E. Schwartz, L.~Martin, E.~Flurin, C.~Aron, M.~Kulkarni, H.~E. Tureci, and
  I.~Siddiqi, ``Stabilizing entanglement via symmetry-selective bath
  engineering in superconducting qubits,'' \emph{Phys. Rev. Lett.}, vol. 116,
  p. 240503, 2016.

\bibitem{Wrachtrup2016}
D.~D.~B. Rao, S.~Yang, and J.~Wrachtrup, ``Dissipative entanglement of
  solid-state spins in diamond,'' 2016, eprint arXiv:1609.00622.

\bibitem{Ticozzi2012}
F.~Ticozzi and L.~Viola, ``Stabilizing entangled states with quasi-local
  quantum dynamical semigroups,'' \emph{Phil. Trans. R. Soc. London A}, vol.
  370, p. 5259, 2012.

\bibitem{TicozziQIC2014}
------, ``Steady-state entanglement by engineered quasi-local {M}arkovian
  dissipation: {H}amiltonian-assisted and conditional stabilization,''
  \emph{Quantum Inf. Comput.}, vol.~14, p. 265, 2014.

\bibitem{Johnson2016}
P.~D. Johnson, F.~Ticozzi, and L.~Viola, ``General fixed points of quasi-local
  frustration-free quantum semigroups: from invariance to stabilization,''
  \emph{Quantum Inf. Comput.}, vol.~16, p. 0657, 2016.

\bibitem{heisenberglyap}
Y.~Pan, Z.~M. H.~Amini, J.~Gough, V.~Ugrinovskii, and M.~R. James, ``Heisenberg
  picture approach to the stability of quantum {M}arkov systems,'' \emph{J.
  Math. Phys.}, vol.~55, 2014.

\bibitem{Ticozzi-alternating}
F.~Ticozzi, L.~Zuccato, P.~D. Johnson, and L.~Viola, ``Alternating projections
  and discrete-time stabilization of quantum states,'' 2016, eprint
  arXiv:1612.05554.

\bibitem{johnson-FTS}
P.~D. Johnson, F.~Ticozzi, and L.~Viola, ``Exact stabilization of entangled
  states in finite time by dissipative quantum circuits,'' 2017, submitted, arXiv:1703.06183.

\bibitem{Blatt2013}
P.~Schindler, M.~Muller, D.~Nigg, J.~T. Barreiro, E.~A. Martinez, M.~Hennrich,
  T.~Monz, S.~Diehl, P.~Zoller, and R.~Blatt, ``Quantum simulation of dynamical
  maps with trapped ions,'' \emph{Nature Phys.}, vol.~9, p. 361, 2013.

\bibitem{Liang2016}
C.~Shen, K.~Noh, V.~V. Albert, S.~Krastanov, M.~H. Devoret, R.~J. Schoelkopf,
  S.~M. Girvin, and L.~Jiang, ``Quantum channel construction with circuit
  quantum electrodynamics,'' 2016, eprint arXiv:1611.03463.

\bibitem{hypergraph}
A.~Bretto, \emph{Hypergraph Theory: An Introduction}.\hskip 1em plus 0.5em
  minus 0.4em\relax Springer, Cham, 2013.

\bibitem{kraus}
K.~Kraus, \emph{States, Effects, and Operations: Fundamental Notions of Quantum
  Theory}, ser. Lecture notes in Physics.\hskip 1em plus 0.5em minus
  0.4em\relax Springer-Verlag, Berlin, 1983.

\bibitem{Bravyi2005}
S.~Bravyi and M.~Vyalyi, ``Commutative version of the local hamiltonian problem
  and common eigenspace problem,'' \emph{Quantum Info. Comput.}, vol.~5, p.
  187, 2005.

\end{thebibliography}

\end{document}